# Leveraging Machine Learning Techniques in Intrusion Detection Systems for Internet of Things

Saeid Jamshidi[1], Amin Nikanjam[1], Nafi Kawser Wazed[1], Foutse Khomh[1]

[1]SWAT Laboratory, Polytechnique Montréal, Montréal, Canada
jamshidi.saeid@polymtl.ca

## Contents

# 1 Introduction

Securing interconnected devices has become a critical concern in the era of pervasive Internet of Things (IoT) technology. Such devices, from household items to complex industrial systems, often handle sensitive data, making them prime targets for cyber threats. This book chapter explores the integration of Machine Learning (ML) techniques into Intrusion Detection Systems (IDS) to enhance the security of IoT networks. Traditional IDS struggle to cope with the vast and varied data generated by IoT devices and the sophisticated nature of modern cyber threats. By leveraging the adaptive and predictive capabilities of ML-based IDS, we can significantly improve the detection and mitigation of intrusions.
The book chapter examines various ML-based IDS techniques' strengths, weaknesses, and applicability to IoT security, emphasizing their potential to detect and respond to complex and evolving threats. Furthermore, we overview critical challenges and highlight ethical considerations and privacy concerns in deploying these advanced IDS technologies. Through this analysis, we aim to provide a robust framework for developing effective, adaptive, and intelligent IDS solutions, ensuring the security and integrity of IoT systems. Finally, we explore the potential contributions of Large Language Models (LLMs) in further enhancing IDS capabilities.
IoT connects many devices, from simple home appliances to complex industrial machinery, marking a significant technological shift. While this interconnection does improve accessibility, it also poses numerous security risks. Hackers target these devices because they frequently gather, process, and transfer sensitive data [1]. The implications of such vulnerabilities extend beyond mere data breaches, posing significant risks to personal privacy and corporate integrity and even compromising national security [2] [3] [4]. Therefore,

IDS is becoming essential for protecting IoT networks. As a digital defense mechanism for IoT, IDS can detect suspicious activities or unauthorized network intrusions. However, due to the complexity of modern cyber threats and the vast volume and diversity of data generated by IoT devices, more advanced solutions are necessary [5].

ML techniques have demonstrated significant potential in fortifying the capabilities of IDS. Providing advanced threat detection capabilities, ML techniques go beyond the limits of traditional rule-based IDS [6]. In the complex and ever-changing world of IoT security, their learning capacity makes them indispensable in keeping up with the ever-changing cyber threats.

This book chapter aims to explore the integration of ML into IDS in the context of IoT, offering an in-depth analysis of the current challenges in this critical field. From the foundational concepts of IDS and ML to cutting-edge research and ethical considerations, we will navigate the complexities of securing the interconnected world. As we investigate the technicalities, case studies, and practical challenges, this chapter underscores the immense potential of ML in revolutionizing IDS for IoT, balancing performance with privacy and ethical considerations, including the capabilities of LLMs in further enhancing IDS. In embarking on this exploration, we are not just addressing the technical audience but also policymakers, security professionals, and practitioners keen on understanding the future of cybersecurity in the IoT era. Through this chapter, we aim to shed light on the pivotal role of ML in crafting robust, adaptive, and intelligent security measures for the IoT.

The remainder of this book chapter is structured as follows: Section 1 introduces the background and motivation for integrating ML and Deep learning (DL) into IDS for IoT. Section 2 provides a detailed classification of IDS methodologies, including placement strategies and intrusion types, and discusses the challenges associated with these systems. Section 3 explores various ML techniques employed in IDS for IoT, highlighting their applications, strengths, and weaknesses. Section 4 considers DL techniques and their specific roles in enhancing IDS capabilities for IoT security. Section 5 reviews significant technical challenges and opportunities in deploying ML and DL-based IDS. Section 6 discusses strategies for overcoming these challenges and highlights future research directions. Finally, Section 7 concludes this chapter's key findings and contributions.

Table 1: List of Abbreviations and Descriptions

| **Abbreviation** | **Description** |
|---|---|
| IoT | Internet of Things |
| IDS | Intrusion Detection Systems |
| AI | Artificial Intelligence |
| ML | Machine Learning |
| DL | Deep Learning |
| SVM | Support Vector Machine |
| LSSVM | Least Squares Support Vector Machine |
| NB | Naive Bayes |
| KNN | K-Nearest Neighbor |
| DT | Decision Tree |
| RF | Random Forest |
| LSTM | Long Short-Term Memory |
| CNN | Convolutional Neural Network |

*Continued on next page*

| Abbreviation | Description |
|---|---|
| AE | Autoencoders |
| RNN | Recurrent Neural Network |
| DBN | Deep Belief Network |
| GAN | Generative Adversarial Network |
| PSO | Particle Swarm Optimization |
| ACO | Ant Colony Optimization |
| PCA | Principal Component Analysis |
| SMOTE | Synthetic Minority Over-sampling Technique |
| RL | Reinforcement Learning |
| LLMs | Large Language Models |
| LLaMA | Large Language Model Meta AI |
| CFL | Class-wise Focal Loss |
| RBM | Restricted Boltzmann Machine |
| ICS | Industrial Control Systems |
| FR | Feature Reduction |
| RBF | Radial Basis Function |
| MMBO | Modified Monarch Butterfly Optimization |
| BRO | Battle Royale Optimization |
| BOA | Butterfly Optimization Algorithm |
| CBOA | Chaotic Butterfly Optimization Algorithm |
| VAE | Variational Autoencoder |
| ASRNN | Attention Segmental Recurrent Neural Network |
| AAE | Adversarial Autoencoders |
| IIoT | Industrial Internet of Things |
| WSN | Wireless Sensor Networks |
| DDoS | Distributed Denial of Service |
| DoS | Denial of Service |
| U2R | User-to-Root |
| R2L | Remote-to-Local |
| Probe | Probe Attacks |
| MLP | Multi-Layer Perceptron |
| GA | Genetic Algorithm |
| ASO | Atom Search Optimization |
| PDAE | Parallel Deep Auto-Encoders |
| MITM | Man-In-The-Middle |
| EO | Equilibrium Optimization |
| GWO | Grey Wolf Optimization |
| CFLVAE | Class-wise Focal Loss Variational Autoencoder |

# 2 IDS in IoT

The classification of IDS is organized according to several characteristics, including the IDS's placement strategy, the IDS's analysis strategy, the type of intrusions, and the attack detection method [7]. This classification for IoT-specific IDS is illustrated in Figure 1.

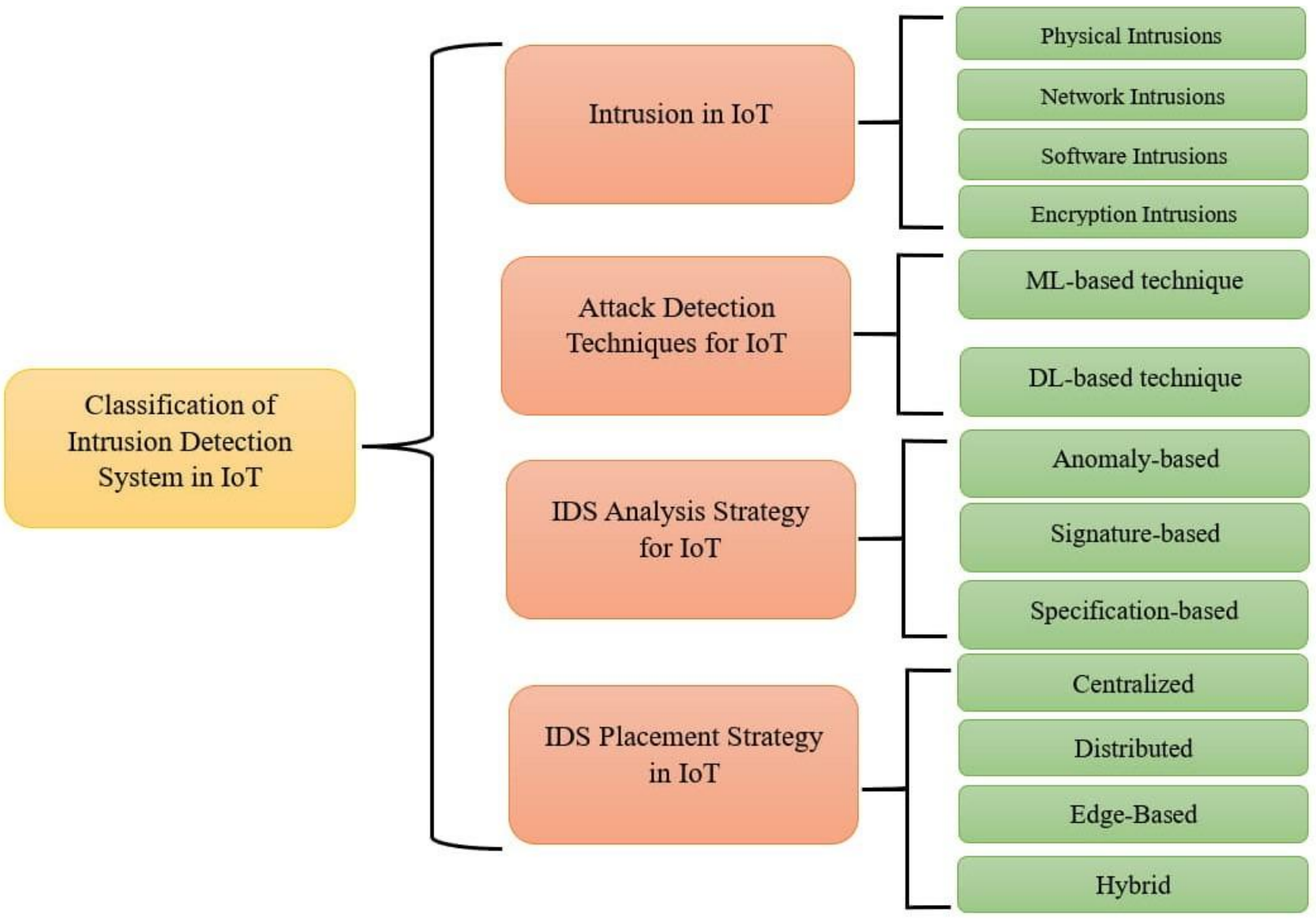

Figure 1: Classification of IDS within IoT.

## 2.1 IDS placement strategy in IoT

The IDS placement strategy in IoT [8] [9] refers to the methods for determining the optimal locations in the IoT infrastructure for the deployment of IDS with various architectural choices (e.g., centralized and distributed) [10] [11]. Additionally, it involves assessing the specific needs and constraints of the IoT (e.g., scalability, resource availability, etc) to ensure network coverage, efficient data analysis, and minimal impact on system performance. The IDS placement strategy is foundational in crafting a robust and responsive security posture that safeguards the IoT from intrusions while maintaining the operational integrity and performance of the network [7] [12] [13].

### 2.1.1 Centralized Placement:

In a centralized IDS, data from various nodes in the network is transmitted to a central location where the detection process is conducted. This model simplifies management and updates to the IDS but makes latency and requires significant bandwidth for data transmission leading to bottlenecks in large-scale IoT [14].

### 2.1.2 Distributed Placement:

Distributed IDS consists of situating detection mechanisms across multiple locations in IoT. This strategy increases the scalability and resilience of the IDS, reducing latency by processing data closer to its source. However, it increases the complexity of managing and synchronizing multiple IDS nodes and requires sophisticated coordination mechanisms [15].

### 2.1.3 Edge-Based IDS Deployment

With the advent of edge computing, placing IDS functionalities at the network's edge, closer to IoT devices, has emerged as an effective strategy. This approach leverages the computational capabilities of edge nodes to perform real-time data analysis and IDS, minimizing latency and reducing the load on central servers or cloud-based systems. Edge-based IDS deployment is especially advantageous for time-sensitive applications, offering rapid response capabilities to mitigate threats promptly [16] [17] [18].

### 2.1.4 Hybrid IDS Placement Strategies

Hybrid strategies combine elements of centralized, distributed, and edge-based approaches to leverage the strengths of each. For instance, lightweight detection algorithms are deployed on edge nodes for immediate threat identification, while analyses requiring extensive computational resources are conducted on centralized servers. This balanced approach optimizes the detection capabilities and resource utilization in IoT [19].

## 2.2 Intrusion in IoT

Intrusions are classified into several categories, each with distinct characteristics for IoT security:

### 2.2.1 Physical Intrusions

Physical intrusion occurs when an unauthorized party gains direct physical access to IoT devices. This could mean tampering with the actual hardware of the device (e.g., sensors, cameras, or storage units). The widespread deployment of IoT devices in locations that can often be remote, exposed, or not secured (e.g., outdoor environments) makes them particularly susceptible to physical attacks. Attackers may extract data directly from the device, install malicious firmware, or even replace it with a compromised device, creating a gateway for further infiltration or disruption. Physical security controls are a fundamental countermeasure, yet they can be challenging to enforce consistently, especially in large-scale or geographically dispersed IoT deployments [20] [21].

### 2.2.2 Network Intrusions

Network intrusions are cyber threats aimed at the communication links and protocols IoT devices use to connect to the internet or other devices. Attackers exploit vulnerabilities within the network infrastructure to gain unauthorized access. Standard methods include sniffing data packets to extract sensitive information and performing Man-In-the-Middle (MITM) attacks to alter communications. The intrusions can have widespread repercussions, impacting not just single devices but potentially entire networks of connected IoT devices. Mitigation strategies involve using secure communication protocols, regular network monitoring, and network security appliances that detect and prevent unauthorized access [22] [23].

### 2.2.3 Software Intrusions

Software intrusions take advantage of vulnerabilities within the software components of IoT. These vulnerabilities are present in the device's firmware and third-party applica-tions running on the device. Common software intrusions include malware infection, which is used to control the device remotely or extract data, and ransomware, which locks users out of their devices until a ransom is paid. These cyber threats are especially concerning because they impact millions of devices simultaneously if a common vulnerability is exploited. Regular software updates and rigorous software testing are critical to safeguarding against such intrusions [24] [25].

### 2.2.4 Encryption Intrusions

Encryption intrusions are sophisticated cyber threats that target the cryptography mechanisms protecting the data exchanged between IoT devices and the systems they communicate with. Hackers attempt to decrypt secure communications through various means, including exploiting weaknesses in the cryptography algorithms or leveraging flaws in their implementation. They also attempt side-channel attacks, which infer sensitive information from the physical implementation of the cryptography system (e.g., energy consumption). These cyber threats are especially challenging to detect as they leave no obvious traces and require high expertise to execute successfully. Ensuring the use of strong, well-implemented encryption algorithms of the latest advances in cryptography is essential to defending against these intrusions [26] [27].

## 2.3 IDS Analysis Strategies for IoT

IDS are classified into four types based on the analysis strategy adopted for detecting intrusions: anomaly-based IDS, signature-based IDS, specification-based IDS, and hybrid IDS. This section discusses IDS techniques developed for IoT.

### 2.3.1 Anomaly-based Detection in IoT

Anomaly-based detection systems are especially advantageous in IoT characterized by dynamic and unpredictable patterns of interaction [28] [29]. These systems monitor network traffic, setting a baseline for regular operation. Any deviation from this established norm can then be flagged for further investigation. The strength of anomaly-based detection lies in its ability to potentially identify new and unforeseen threats, making it invaluable in scenarios where IoT devices exhibit a wide range of behaviors and where new devices may frequently connect or disconnect. However, the flexibility of this approach comes with the challenge of distinguishing between genuine threats and benign anomalies, necessitating sophisticated algorithms to reduce false positives [30] [31].

### 2.3.2 Signature-based Detection in IoT

Signature-based detection is highly effective in contexts with stable and predictable digital ecosystems, making it a fundamental component of IDS [32]. It operates by comparing observed data against a database of known threat signatures, unique sets of data, or attributes that are known to be malicious. This method is suited for static IoT where devices execute limited operations and network traffic patterns are consistent. While highly effective against recognized threats, its primary limitation is the inability to detect novel cyber threats, which makes regular signature database updates a critical part of maintenance for IoT [33] [34].

### 2.3.3 Specification-based Detection in IoT

Specification-based detection is predicated on establishing formal models of correct device behavior within an IoT. This method is especially salient in industrial IoT, where devices operate under stringent protocols, and any deviation from these protocols signals a security breach [35]. Specification-based systems offer the precision necessary for environments where operational parameters are clearly defined and immediately categorized as intrusions. The challenge with this approach is the considerable effort required to define and maintain accurate specifications, especially in complex systems with diverse and evolving device functionalities [36] [37].

### 2.3.4 Hybrid Approaches in IoT

Hybrid IDS approaches represent a confluence of the various analysis techniques, tailored to leverage their respective strengths to fortify IoT security. By integrating anomaly, signature, and specification-based strategies, hybrid systems offer a robust defense against a wide array of known and unknown threats [38] [39]. This multifaceted method is especially beneficial in complex IoT that witness a convergence of public, private, and industrial devices, each with unique security requirements. A hybrid approach provides a nuanced security posture that adapts to the intricate threatscape of IoT. However, its implementation is complicated, requiring careful orchestration to ensure the seamless operation of the combined methodologies [40].

# 3 ML-based IDS for IoT

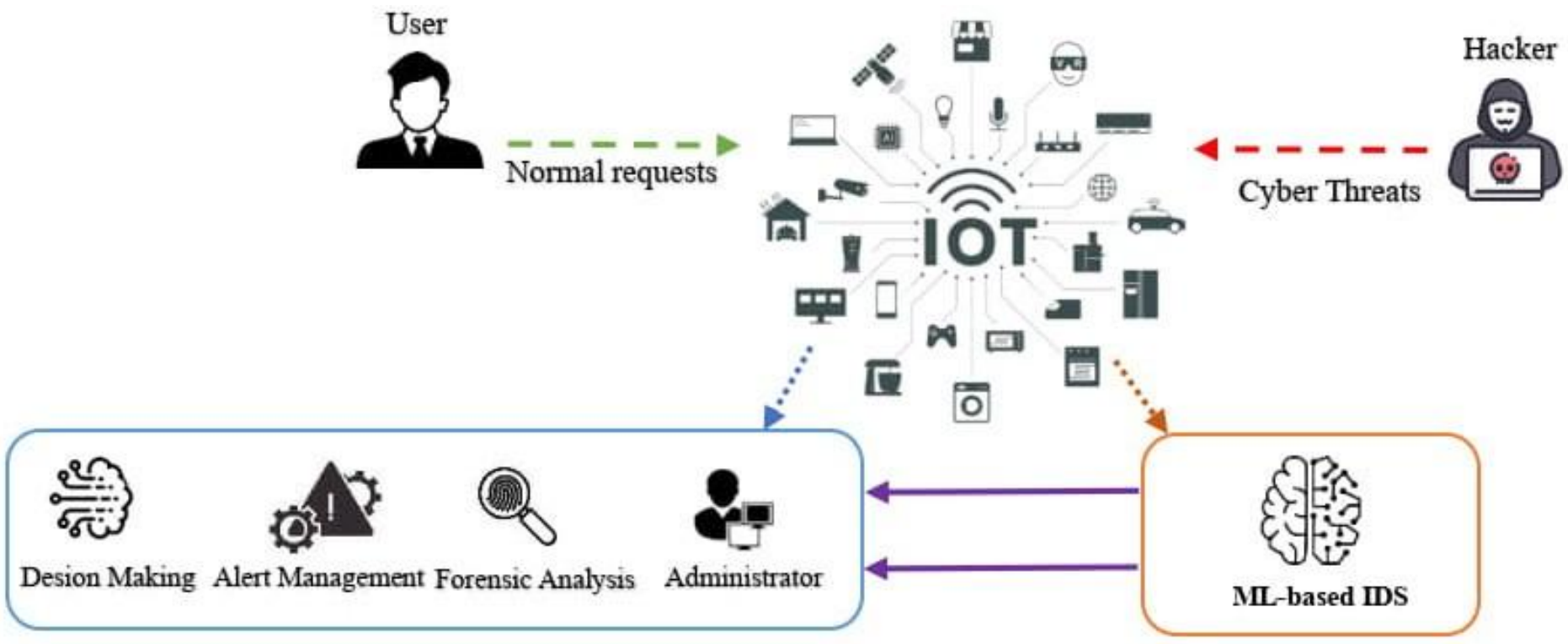

Figure 2: A sample depiction of the cyber threat detection environment based on ML.

ML, a branch of Artificial Intelligence (AI), utilizes various techniques to analyze data to uncover patterns, thereby facilitating predictive insights. Drawing from a rich tapestry of disciplines, including mathematics and computer science, ML has catalyzed transformations across numerous sectors by addressing complex problems with innovative solutions [41]. The influence of ML spans a wide array of applications, from enhancing facial recognition technologies for social media interactions to advancing capabilities in optical character recognition, recommendation systems, and autonomous vehicles.
Within the vast and complex landscape of the IoT, ML significantly bolsters the efficacy of IDS. It enables IDS to navigate and interpret the massive, heterogeneous data streams produced by IoT devices, a critical capability for differentiating between standard operations and potential security threats. ML's application in this domain is adeptly facilitated by supervised, unsupervised, and Reinforcement Learning (RL) techniques [42]. The intricate dynamics of ML's integration into IDS in the IoT ecosystem, including the distinctions between its various sub-domains, are visually represented in Figure 3.
In the specific context of IDS tailored for IoT, the deployment of ML is pivotal for developing highly accurate systems. In the rest of this chapter, we will explore the foremost ML methodologies, highlighting their particular relevance and application in crafting IDS. These solutions are adept at accurately identifying deviations from typical behavioral patterns, thereby safeguarding the security and integrity of IoT infrastructures. This exploration will include a detailed examination of Figure 2, which schematically illustrates the role of ML in the IoT.

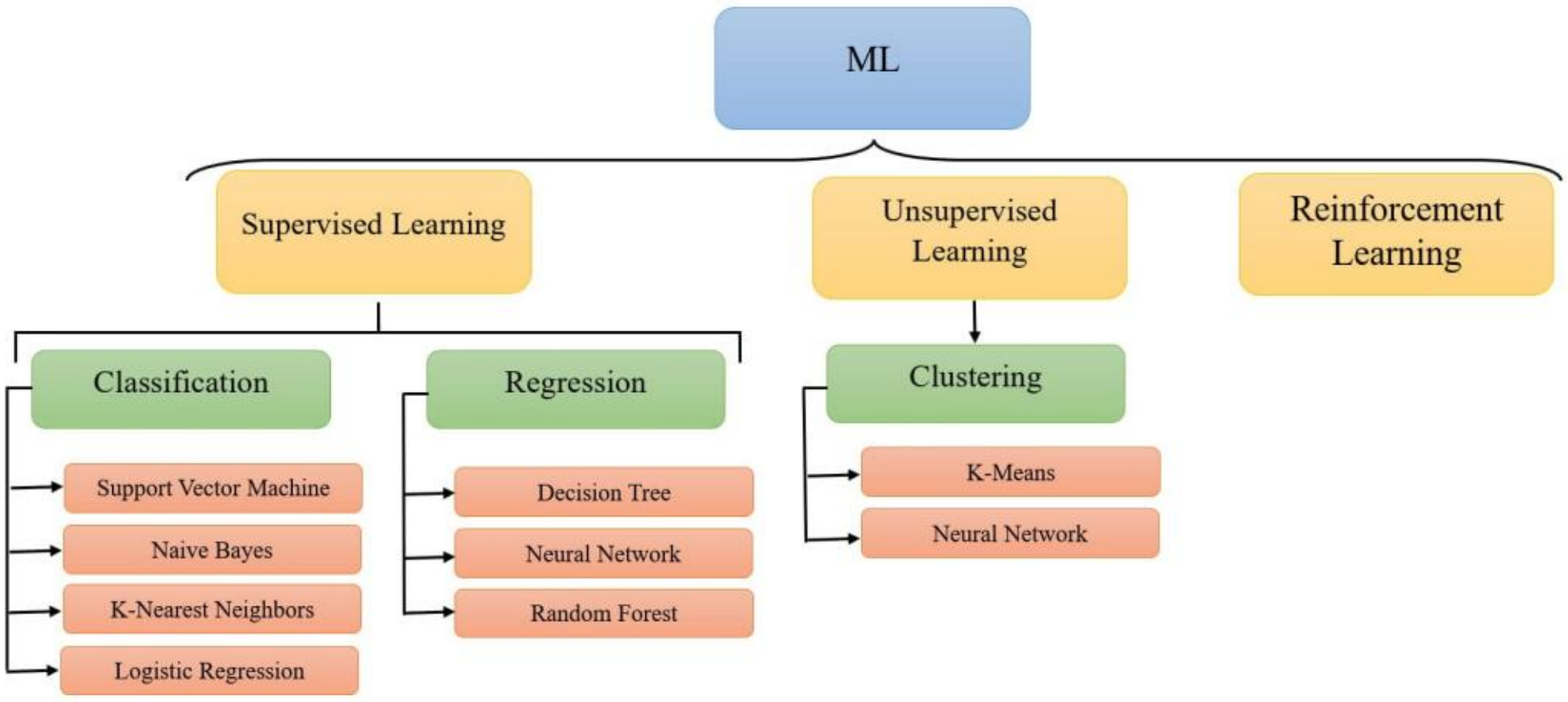

Figure 3: ML methods.

### 3.0.1 The Support Vector Machine (SVM)

SVM is a sophisticated supervised learning algorithm extensively utilized for classification, regression, and outlier detection. Their core strategy involves generating a hyperplane within a high-dimensional feature space to separate data linearly, maximizing the margin between different class data points. Originally designed for binary classification, SVM's versatility extends to multi-class scenarios, proving effective in handling nonlinear data.
Parveen Akhtar et al. [43] presented a novel approach to enhancing the security of IoT networks. The proposed system employs the Least Squares Support Vector Machine (LSSVM) technique to identify potential intrusions accurately. This approach involves preprocessing data through normalization, discretization, and feature selection, which are crucial for preparing the dataset for model training. The IoTID20 dataset, comprising data from smart home environments, serves as the testbed for this study. The findings indicate that the LSSVM-based model outperforms traditional methods such as SVM and RFs, achieving an accuracy of 97.7%. This research underscores the importance of advanced ML techniques in developing robust IDS capable of addressing the unique security challenges of IoT networks.
Ahmed Abdullah Alqarni et al. [44] proposed a novel IDS that combines SVM with Ant Colony Optimization (ACO is a probabilistic technique for solving computational problems that involve finding good paths through graphs. Inspired by the behavior of ants searching for the shortest path to food, ACO utilizes a colony of artificial ants that simulate the deposition of pheromones on a graph to form a path between points. Over time, the path with the strongest pheromone trail provides the optimal solution. ACO is particularly effective in optimizing network-based problems and feature selection tasks.) to enhance cybersecurity. The study addresses the challenge of detecting malicious activities within network traffic by employing ACO to reduce the dimensionality of large datasets, specifically KDD-Cup99 and NSL-KDD, thereby selecting the most significant features for SVM classification. The empirical findings demonstrate that the hybrid SVM-ACO model achieves superior accuracy in detecting various types of network intrusions, including Denial-of-Service (DoS), User-to-Root (U2R), and Remote-to-Local (R2L) attacks.
Almaiah et al. [45] explored the efficacy of integrating Principal Component Analysis (PCA) with SVM classifiers to enhance IDS. The study evaluates different SVM kernel functions: linear, polynomial, Gaussian Radial Basis Function (RBF), and sigmoid, using the KDD-Cup99 and UNSW-NB15 datasets. The findings reveal that the Gaussian RBF kernel consistently outperforms other kernels in terms of detection accuracy such as 99.11% for KDD Cup'99 and 93.94% for UNSW-NB15.

### 3.0.2 Naive-Bayes (NB)

The NB classifier presupposes the independence of each feature within a class for making predictions. It calculates the probabilities for each class given a specific instance, selecting the class with the highest probability for its prediction. The NB in IDS has been explored in numerous studies.
Kevric et al. [46] crafted a hybrid classifier model that merges random tree and NBTree algorithms tailored for network IDS, achieving an 89.24% accuracy on the NSL-KDD dataset. Their findings reveal that the simple aggregation of top-performing classifiers doesn't necessarily yield the best collective performance.
To address diverse attack vectors, Çavuşoğlu [47] devised a hybrid layered IDS employing a blend of ML techniques, validated using the NSL-KDD dataset. The methodology underscored accuracy and a reduced false positive rate across various attack scenarios. Gu and Lu [48] advanced this further by utilizing SVM and NB feature embedding to develop an effective IDS framework, showcasing the potential of marrying different ML strategies to enhance network security.
Vishwakarma et al. [49] proposed a two-phase ML-based IDS specifically for IoT. The initial phase employs different versions of the NB classifier, supplemented by majority voting for the final classification. In contrast, the subsequent phase uses an unsupervised elliptic envelope to scrutinize initially further deemed normal data. Evaluated against NSL-KDD, UNSW-NB15, and CIC-IDS2017 datasets, their method achieves a 97% accuracy on the NSL-KDD dataset and underscores significant potential in enhancing IoT security.
Jeevaraj et al. [50] introduced an NB ML algorithm-based feature selection model to fortify IDS within Wireless Sensor Networks (WSNs). Addressing the security challenges posed by the proliferation of WSNs, their methodology focuses on streamlining feature selection to augment the IDS. By optimizing IDS using a minimal set of features, the study achieved a remarkable prediction accuracy of nearly 95.8%, a precision level of 95%, and an area under the curve of 0.98%.

### 3.0.3 K-Nearest Neighbor (KNN)

KNN algorithm is recognized for its adaptability across both classification and regression tasks, with a strong suitability for classification. As a lazy learning algorithm, KNN conservatively maintains all training data, leveraging this dataset to assess similarities between known data points and new queries. The algorithm classifies each test instance based on the distance to its nearest neighbors, attributing the instance to the class with the closest proximity. Despite its effectiveness, KNN's reliance on extensive distance computations across large datasets is computationally demanding.

Guo et al. [51] innovatively combined anomaly and misuse detection components in a two-tier hybrid IDS model. An initial anomaly detection layer preliminarily filters data, which KNN then refines to minimize false positives and negatives in the subsequent stage. Demonstrated on the KDD-Cup99 and Kyoto University Benchmark datasets, this model excelled in network anomaly detection with minimal false positive rates.

Saleh et al. [52] tackled multi-class classification challenges within IDS by devising a hybrid system underpinned by a tripartite strategy. This system utilized NB feature selection for dimensionality reduction, a one-class SVM for outlier rejection, and proximity KNN for direct attack detection. Evaluated across multiple datasets, including the KDD-Cup99 and NSL-KDD, the system demonstrated prompt attack detection capabilities, underscoring its potential for real-time IDS applications and marking a forward leap in IDS methodology.

Mohy-Eddie et al. [53] developed a novel IDS leveraging KNN and election-based feature selection to fortify IoT security. This model enhances IoT protection by employing anomaly-IDS and selecting key features through PCA, Genetic Algorithm (GA), and univariate statistical tests. This strategic feature selection significantly reduced training durations while maintaining high accuracy on the Bot-IoT dataset, showcasing the model's efficiency and effectiveness.

### 3.0.4 Decision Tree (DT)

DT is pivotal for IDS, excelling in classifying network events into normal activities or potential threats through a clear, hierarchical decision-making process. Their structure simplifies the complex analysis of network data, making DTs especially effective for IDS by offering straightforward interpretability and detailed insights into each classification decision. The classification and regression tree model, a DT variant, enhances IDS with its binary trees, streamlining decision paths for rapid and accurate threat detection. This method's capacity for clear, rule-based decision-making aligns seamlessly with the needs of IDS, ensuring security analysts can quickly understand and respond to identified threats.

Mehta et al. [54] developed an innovative IDS tailored for the controller area network protocol, which is commonly used in the automotive and aerospace industries. They introduce the ML-based approach, utilizing DT ensembles ( e.g., Adaboost, Gradient boosting) and Feature Reduction (FR) to detect anomalies in traffic effectively. This method outperforms traditional techniques by achieving near-perfect detection rates with an area under curve score of up to 0.99%, demonstrating its robustness and reliability.

Guezzaz et al. [55] presented an enhanced IDS that utilizes a DT classifier improved with data quality techniques for more reliable network security. The approach involves preprocessing network data, selecting relevant features using entropy decision techniques, and then training a DT to distinguish between normal and abnormal network activities. Their experiments on the NSL-KDD and CICIDS2017 datasets demonstrate that this method achieves high accuracy and detection rates, specifically 99.42% and 98.2% detection rate on the NSL-KDD dataset, and 98.8% and 97.3% detection rate on the CICIDS2017 dataset.

Umar et al. [56] presented a hybrid IDS modeling approach combining a DT-based wrapper feature selection method with various ML algorithms to build efficient and accurate IDS models. They utilized the UNSW-NB15 dataset and compared their approach with baseline models that use complete feature sets. Their findings demonstrate that this method improves the intrusions' detection rate and reduces the computational time needed for model training. Specifically, they achieved a detection rate of 97.95% with reduced feature sets, demonstrating the effectiveness of their approach in handling large and complex data while maintaining high performance in identifying IoT threats.

### 3.0.5 Random Forest (RF)

RF is an ensemble learning method that builds a forest from multiple DTs. By averaging the predictions from each tree, RF typically yields more accurate predictions than a single DT. The ensemble approach of RF means that as the number of trees increases, the overall model becomes more stable and robust. This method effectively improves predictive performance by combining the outcomes of various trees within the forest.

Ajdani et al. [57] introduced an enhanced IDS method that leverages the RF algorithm in combination with the Particle Swarm Optimization ( PSO is a computational method that optimizes a problem by iteratively improving a candidate solution about a given measure of quality. It simulates the social behavior of birds within a flock or fish school. PSO is initialized with random solutions and searches for optima by updating generations. In every iteration, each particle adjusts its position based on its own experience and the experience of neighboring particles, making it excellent for optimizing complex spaces and enhancing algorithm performance.) algorithm. RF, an ensemble learning method, improves predictive accuracy by averaging results from multiple DTs, increasing robustness as more trees are added. Each tree is trained on a bootstrapped subset of data, and a subset of features is selected randomly at each split, enhancing the model's generalizability. The PSO algorithm optimizes this process by guiding the selection of features and configurations, improving detection rates with greater efficiency. Their approach demonstrates significant performance improvements, with IDS accuracy increased to 97%, highlighting the combined strength of RF and PSO in reducing false positives and enhancing speed and accuracy in identifying network threats.

Zhang et al. [58] proposed an enhanced approach to IDs that utilizes a novel three-branch RF algorithm. This method addresses the challenge of distinguishing important attributes in network data by using decision boundary entropy to assess attribute importance and dividing them into three categories: positive, boundary, and negative domains. By enhancing the RF model with this three-way decision-making process, the approach improves the performance metric of IDS, making it highly effective in identifying and classifying network intrusions.

Balyan et al. [59] presented an advanced IDS combining enhanced GA with PSO and an improved RF method. This hybrid model effectively addresses the challenges of data imbalance and feature selection in IDS datasets. The model achieves high accuracy and detection rates by optimizing the feature selection process using PSO and enhancing the classification performance with RF.

Wu et al. [60] proposed a novel IDS that combines an enhanced RF model with the Synthetic Minority Over-sampling Technique (SMOTE) to address the challenges of data imbalance and improve accuracy. By integrating the K-means clustering with SMOTE, the method effectively balances the dataset, enhancing the minority class representation and enabling the RF to learn more discriminating features. This hybrid approach significantly reduces false positives and improves the model's sensitivity to various attack types.

### 3.0.6 K-means Clustering

The K-means algorithm is an unsupervised ML technique that operates without labeled data. This method identifies clusters in a dataset by grouping similar objects and distinguishing them from dissimilar ones in other clusters. Commonly utilized in pattern matching within time series data, the K-means algorithm groups data points based on their similarities and differences. However, a significant limitation of this algorithm is its ineffectiveness with non-spherical clusters, as it assumes clusters to be roughly spherical.
Mahdieh Maazalahi et al. [61] presented a two-stage hybrid approach for IDS that combines ML and meta-heuristic algorithms. The first stage involves data preparation, where string values are converted to a numeric format and normalized. For feature selection, Atom Search Optimization (ASO is a nature-inspired algorithm that mimics the behavior of atoms in nature. It is used for finding optimal solutions by exploring the search space through interactions among atoms.) and Equilibrium Optimization (EO is a novel optimization algorithm based on the control-volume mass balance model to simulate the dynamic equilibrium state in a closed system. It is utilized for complex problem-solving where finding an optimal balance is necessary.) Algorithms are used to aim for global optimization. In the second stage, attack detection employs K-means clustering and the Firefly Algorithm (the flashing behavior of fireflies inspires FA. It is used in optimization problems where the attractiveness of a solution is proportional to its brightness or fitness). This method, termed ASO-EO-FA-K-means, is evaluated using the NSL-KDD, UNSW-NB15, and KDD-CUP99 datasets, demonstrating superior accuracy and computational efficiency compared to other methods, such as PSO, GA, Grey Wolf Optimization (GWO), and XGBoost. The hybrid method achieves the highest accuracy with 0.99%, 0.99%, and 0.99% for the datasets respectively, proving its effectiveness in IDS.
Noura Alenezi et al. [62] proposed a novel approach to enhance the security of Industrial Internet of Things (IIoT) environments. The authors introduce an intelligent IDS that employs PCA for feature engineering, significantly reducing data dimensionality and improving detection performance. In the classification phase, clustering algorithms, e.g., K-means, classify IIoT traffic as normal or malicious. The model is validated using the X-IIoTID dataset, which includes a variety of IIoT protocols and cyberattack scenarios. The proposed IDS achieved an impressive accuracy rate of 99.79%, outperforming existing methods.
Amith Murthy et al. [63] addressed the security challenges of integrating IoT devices in smart buildings. The paper develops a device-specific traffic classification model using ML to detect attacks (e.g., Distributed Denial of Service (DDoS)) with high accuracy 96%. Their approach involves classifying traffic flows into four coarse-grained types based on traffic sources and directions, extracting simple yet effective features for learning and prediction. This lightweight model, which only requires 32 features, balances accuracy and efficiency, making it suitable for large-scale smart building networks.
This section highlights the efficacy of traditional ML-based IDS for IoT, as summarized in Table 2. SVM, DT, and RF handle complex scenarios with high robustness, while Naive Bayes and K-Means are effective for simpler tasks, ensuring broad applicability across diverse IoT security challenges.

Table 2: Comparison of Traditional ML Techniques in IDS for IoT

| Feature | SVM | NB | DT | RF | K-Means |
|---|---|---|---|---|---|
| **Type of Learning** | Supervised | Supervised | Supervised | Supervised | Unsupervised |
| **Classification Performance** | High | Moderate | High | Very High | Moderate |
| **Speed of Classification** | Moderate | Fast | Fast | Moderate | Moderate |
| **Handling Non-Linear Data** | Good (with kernel tricks) | Poor | Poor | Good | Moderate |
| **Scalability to Large Data** | Moderate | High | High | High | Moderate |
| **Robustness to Noisy Data** | Moderate | Low | Moderate | High | Low |
| **Interpretability** | Low | High | High | Moderate | Moderate |
| **Computational Complexity** | High (especially with non-linear kernels) | Low | Moderate | High (due to multiple DTs) | Moderate |
| **Typical Use Case in IDS** | Complex pattern recognition, binary classes | Baseline anomaly detection | Rule-based analysis, simple conditions | Ensemble of DTs, complex scenarios | Cluster-based anomaly detection |

### 3.0.7 DataSet

Datasets play a critical role in developing and evaluating IDS for the IoT. They provide the necessary data to train, validate, and test the performance of IDS models. Table 3 of commonly used IoT security datasets and the types of attacks they cover.
Datasets are crucial for ML and DL-based systems as they provide the data needed for training, validating, and testing models. They enable feature learning, help in tuning model parameters, and allow for performance evaluation. Diverse and comprehensive datasets ensure robustness and relevance to real-world scenarios, essential for handling various situations and anomalies. They also facilitate benchmarking and standardization, allowing for comparisons across different models. Handling class imbalances and preprocessing data are vital to ensure high-quality input for effective learning, making datasets the cornerstone of successful ML and DL applications.

# 4 DL-based IDS for IoT

This section explores the effectiveness and strategic implementation of DL-based IDs for the IoT. Here, we will examine a variety of DL models, including Long Short-Term Memory (LSTM) networks, Convolutional Neural Networks (CNN), Autoencoders (AE),

Table 3: Overview of attacks covered by each dataset

| Dataset Name | Attacks Covered |
|---|---|
| ISCX 2012 IDS [64] | Brute Force SSH, FTP, Infiltration, HTTP DoS, DDoS, Botnet |
| N-BaIoT [65] | DDoS, DoS, Recon., MITM |
| UNSW-NB-15 [66] | Fuzzers, Analysis, Backdoors, DoS, Exploits, Recon., Worms |
| CICIDS2017 [67] | DoS, DDoS, Heartbleed, Bot, Infiltration, Web, PortScan, FTP/SSH Patator, etc. |
| IoTID20 [68] | Backdoor, Analysis, Fuzzing, DoS, Exploits, Recon., Worms |
| NSL-KDD [69] | DoS, Probe, U2R, R2L, etc. |
| AWID [70] | Flooding, Injection, Impersonation, Misc. |
| CSE-CIC-IDS2018 [71] | Brute Force, DoS, DDoS, SQL Injection, Heartbleed, etc. |
| WUSTL-IIoT-2021 [72] | APTs, MITM, Side-channel, Zero-day, etc. |
| WUSTL-IIoT-2018 [73] | APTs, Evasion, Side-channel |
| MedBIoT [74] | Early-stage malware (e.g., scanning, recon.) |
| TON-IoT [75] | Botnet, Cmd Injection, Malware Propagation |
| DS2OS Trace [76] | Feature Extraction in IoT system |
| Hogzilla [77] | CTU-13 Botnet, Illegitimate Packets |
| SCADA Power Sys. [78] | Stuxnet worm, Iran nuclear machinery attack |
| CIC-DDOS2019 [79] | DDoS (e.g., UDP, HTTP, SYN) |
| BoT-IoT [80] | DDoS, DoS, Scan, Keylogging, Data exfiltration |
| ISOT-CID [81] | IoT attacks, vulnerabilities |
| BoTNeTIoT-L01 [82] | Automated IoT botnet, Traffic anomalies |
| IEC 60 870-5-104 [83] | Energy sector protocol attacks |

Recurrent Neural Networks (RNN), and Deep Belief Networks (DBN), highlighting their roles in enhancing the detection and prevention of security threats in IoT infrastructures. The discussion will underscore how these DL models, by autonomously learning from and adapting to complex data patterns, offer significant improvements over traditional ML techniques in identifying subtle, sophisticated cyber threats. Figure 4 illustrates various DL-based IDS architectures deployed in IoT security, showcasing how these technologies are implemented to enhance system robustness against cyber threats.

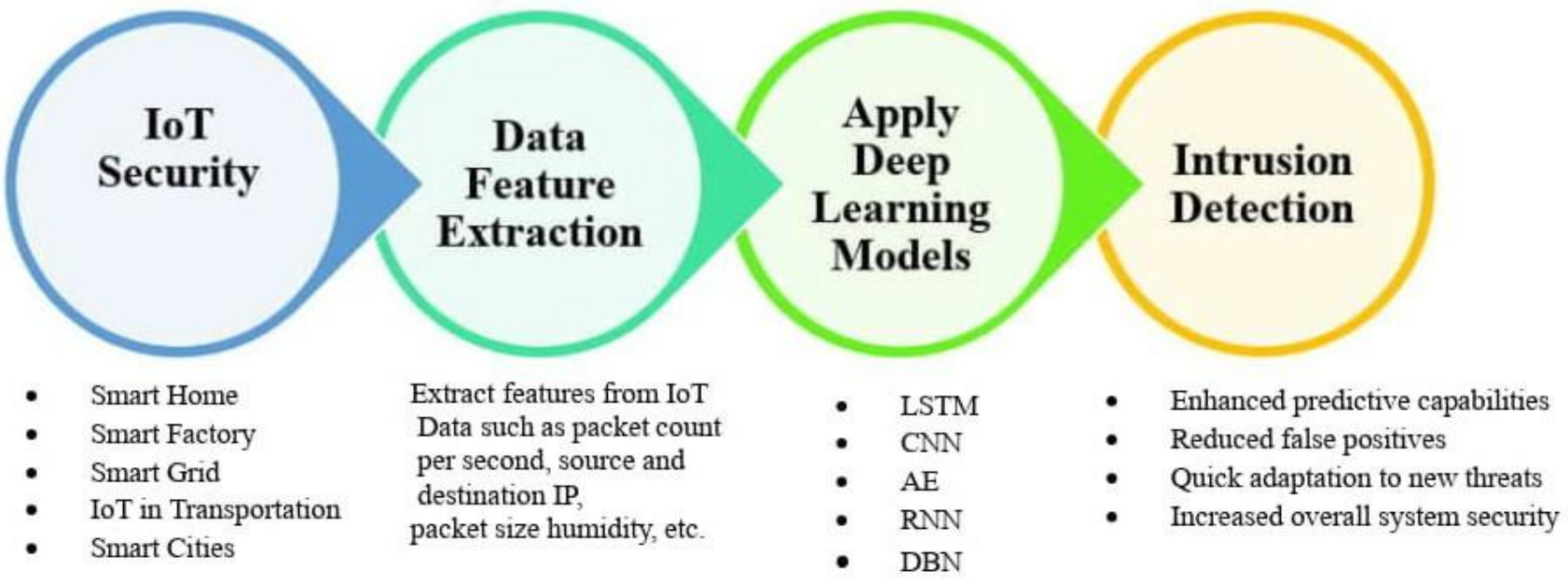

Figure 4: DL-based IDS in IoT security

### 4.0.1 Long Short-Term Memory (LSTM)

LSTM networks are an advanced type of RNN designed to recognize patterns in data sequences, making them ideal for time-series analysis in IoT IDS. By effectively capturing long-range dependencies, LSTMs can identify complex temporal anomalies in network traffic, enhancing the detection of sophisticated cyber threats. Their unique architecture, with input, forget, and output gates, allows them to remember important information over extended intervals and forget irrelevant details. This capability makes LSTMs especially effective for monitoring and securing IoT devices against persistent, evolving threats.
Azumah et al. [84] introduced a novel IDS approach for IoT networks in smart homes using the LSTM model. Their research demonstrated that this model significantly improves the accuracy of detecting cyberattacks compared to traditional methods. The LSTM model effectively monitors and identifies various intrusions, providing timely alerts to users. This capability reduces false alarms and ensures homeowners respond quickly to potential threats.
Awad et al. [85] introduced an enhanced version of the LSTM network, termed improved LSTM, to boost the performance of IDS in network security. Their novel approach integrates a Chaotic Butterfly Optimization Algorithm (CBOA), an advanced version of the traditional Butterfly Optimization Algorithm (BOA), which is inspired by butterflies' natural foraging behavior. CBOA incorporates chaotic sequences into the optimization process to avoid local optima and improve the exploration capabilities of the algorithm. This approach enhances the algorithm's efficiency in navigating complex search spaces. It is beneficial for optimizing neural network parameters, such as weights and learning rates, thereby improving convergence speed and overall performance. with PSO to optimize the LSTM weights more efficiently, aiming to reduce the iterations needed for convergence while improving classification accuracy. Their paper demonstrated that the LSTM model outperforms traditional LSTM and DL models in detecting network intrusions. Specifically, LSTM achieved higher accuracy and precision across two public datasets, NSL-KDD and LITNET-2020, used for binary and multi-class classification tasks. The incorporation of hybrid CBOA and PSO allowed the LSTM to minimize false positives and enhance its detection capabilities for various attack types (e.g., Denial of Service (DoS), Probe attack, and User-to-Root (U2R).
Zhou et al. [86] introduced an incremental LSTM model for enhancing IDS in network security. Their method adapts LSTM to detect dynamic changes in network traffic more effectively by incorporating incremental changes. This approach significantly improved the detection rates and reduced false positives on datasets (e.g., UNSW-NB15 and CICIDS2017), outperforming traditional models. For instance, their model achieved up to 98.01% detection accuracy, illustrating its superior ability to identify known and unknown network intrusions.
Imrana et al. [87] introduced a bidirectional LSTM model to improve the detection capabilities of IDS in network security. By processing network data in both forward and backward directions, their LSTM model effectively captures complex dependencies and anomalies. Their findings showed that the LSTM model significantly outperforms traditional LSTM and other models' performance on the NSL-KDD dataset. The model excelled in detecting sophisticated U2R and Remote-to-Loca (R2L) attacks, demonstrating its potential to enhance network security dynamically.
Alimi et al. [88] developed a refined LSTM model to enhance the IDS of DoS attacks in IoT networks. Their approach improved the detection capabilities by adapting the LSTM model to handle dynamic and large-scale IoT traffic better. The LSTM model outperformed traditional methods, demonstrating superior performance on benchmark datasets (e.g., NSL-KDD and CICIDS-2017). For instance, on the CICIDS-2017 dataset, the model achieved an accuracy of 99.22%.

### 4.0.2 Convolutional Neural Network (CNN)

CNN is highly effective for IDS in IoT due to its ability to automatically learn and identify complex patterns in network traffic. Using convolutional and pooling layers, CNNs can process vast volumes of data and detect anomalies without manual feature extraction. This makes them adept at recognizing various cyber threats in IoT, from malware to unauthorized access, enhancing the accuracy and efficiency of IDS. Their application in IoT IDS helps maintain robust security by swiftly identifying and responding to potential intrusions, as illustrated in Figure 5, which depicts a typical CNN architecture.

Aljumah [89] presented a temporal CNN for enhancing IDS in IoT. The CNN with SMOTE handles imbalanced datasets and employs feature engineering techniques attributes (e.g., transformation and reduction). The proposed model is evaluated using the Bot-IoT dataset and compared to traditional ML algorithms (RF and logistic regression) and other DL models (LSTM and CNN). The results demonstrated that CNN achieved a multi-class traffic detection accuracy of 99.99%.

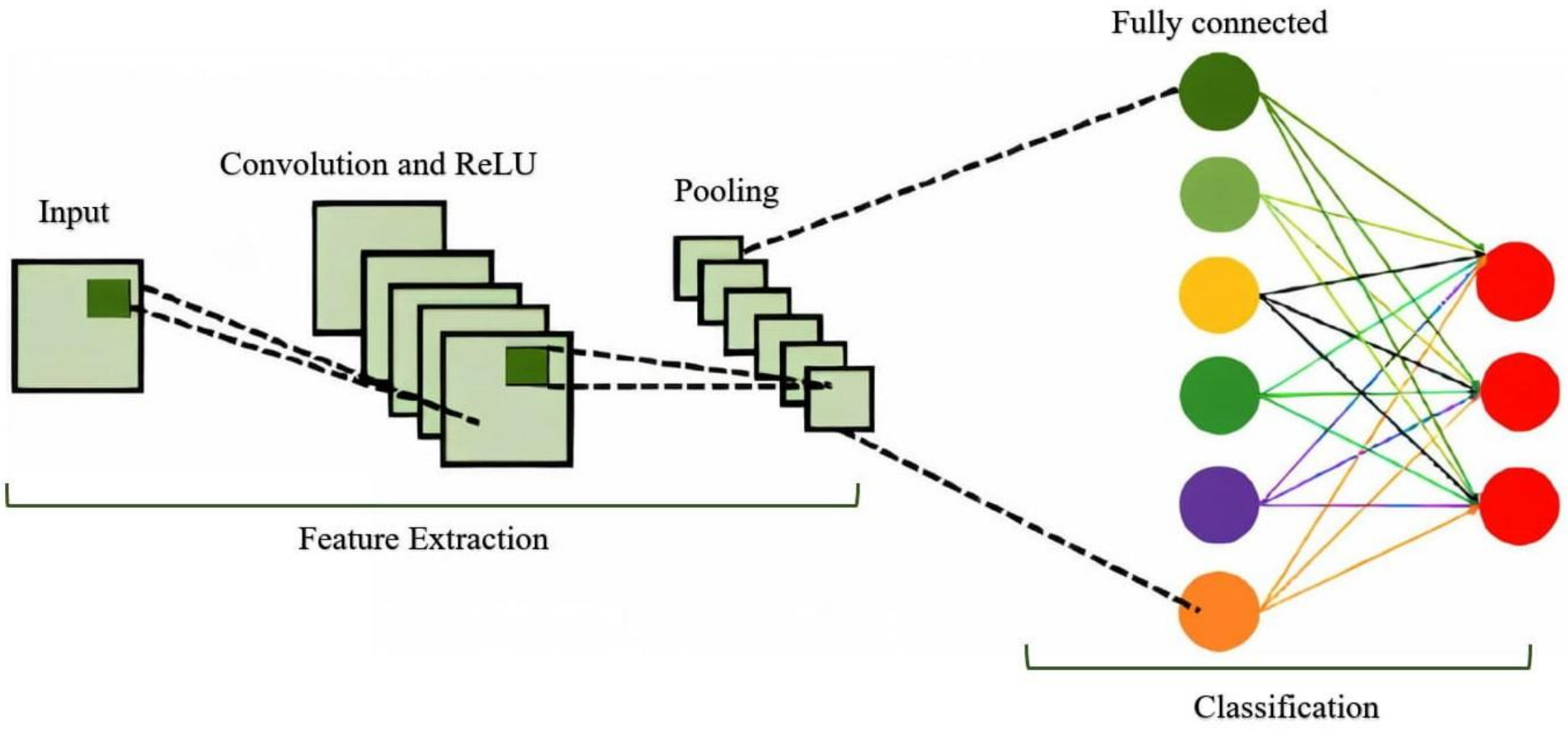

Figure 5: Typical CNN Architecture.

Ho et al. [90] introduced a novel IDS model using a CNN designed to detect a wide range of cyberattacks, including novel and sophisticated threats. Their paper utilized the CICIDS2017 dataset, a source for various network attack vectors, to train and evaluate their model. By optimizing the CNN, they addressed the challenges of class imbalance and feature extraction, which are critical for accurate IDS. The study demonstrated that the CNN model outperformed traditional ML and other neural network models in detecting cyberattacks. Specifically, the model achieved an accuracy of 99.57% on the CICIDS2017 dataset.

Kumar et al. [91] proposed a novel IDS that utilizes the CNN combined with Modified Monarch Butterfly Optimization (MMBO is an enhancement of the original Monarch Butterfly Optimization algorithm. MMBO introduces modifications to the migration operator and integrates new mechanisms for adjusting the control parameters, improving the algorithm's ability to efficiently explore and exploit the search space. This adaptation helps achieve faster convergence and better solution quality, making it suitable for complex optimization tasks such as feature selection in IDS to detect attacks in IoT networks. The model employs min-max normalization for data preprocessing and improved Battle Royale Optimization (BRO), a variant of the Battle Royale Optimization algorithm, inspired by the competitive elimination process in battle royale games. This optimization technique progressively eliminates less effective solutions in a series of competitive rounds, allowing only the strongest solutions to survive. This process enhances the exploration and exploitation capabilities of the algorithm, leading to optimized feature selection in ML models for feature selection, enhancing the classifier's performance. The model was tested on the N-BaIoT and CICIDS-2017 datasets, achieving high accuracy rates of 99.96% and 99.95%, respectively.
Ullah et al. [92] proposed a CNN-based IDS enhanced by feature engineering techniques for IoT networks. The system integrates two convolutional layers and three fully connected layers to optimize performance and computational efficiency, utilizing the IoTID20 dataset for evaluation. The model demonstrated high accuracy and robustness, achieving a binary classification accuracy of 99.84%, multi-class classification accuracy of 98.12%, and subcategory classification accuracy of 77.55%.

### 4.0.3 Autoencoders (AE)

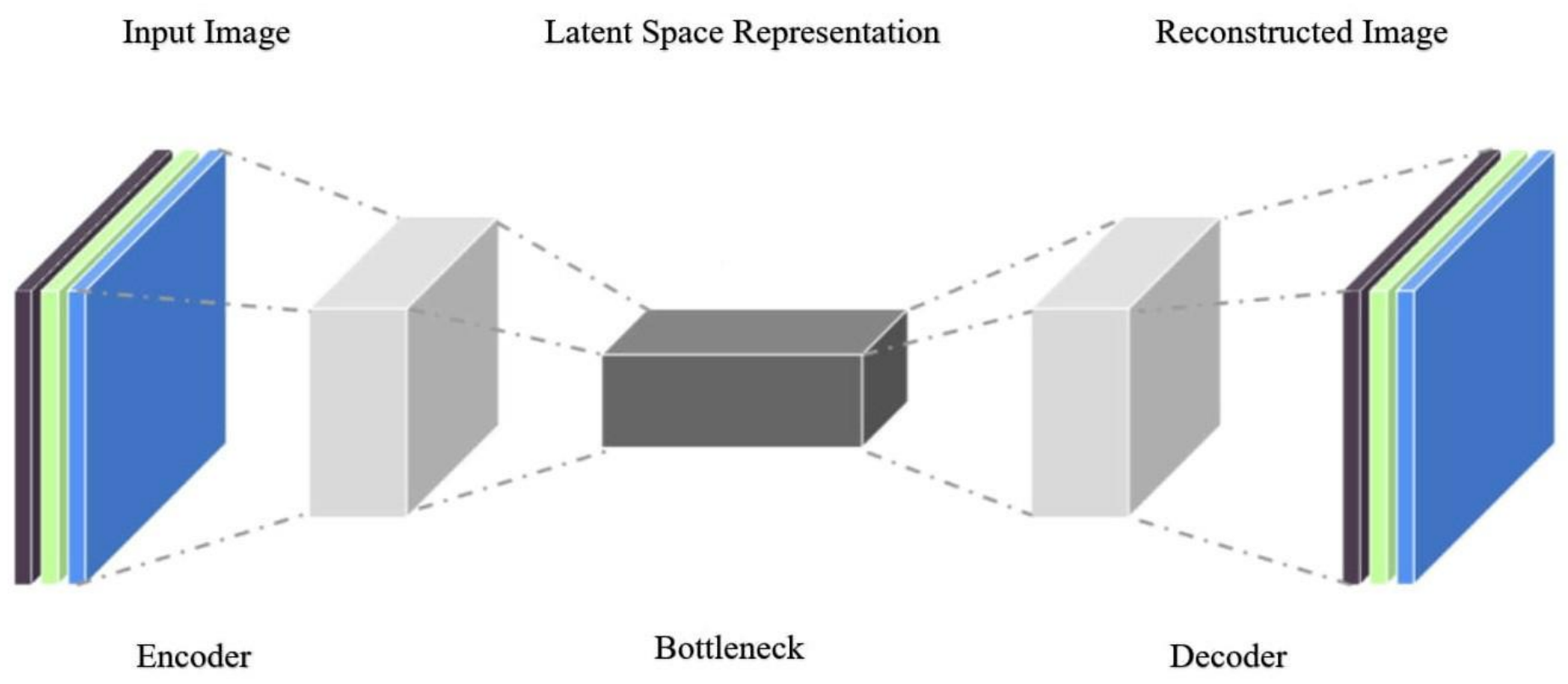

Figure 6: Typical AE Architecture.

AE are unsupervised neural networks used in IoT for IDS to identify anomalies by learning compressed representations of input data. They consist of an encoder that compresses data and a decoder that reconstructs it, minimizing reconstruction error. AEs distinguish normal from anomalous data by exhibiting higher reconstruction errors for atypical patterns, effectively detecting intrusions. Their scalability, unsupervised learning capability, and efficiency in handling large IoT data volumes enhance IDS performance. Implementing AEs involves data collection, preprocessing, training on normal data, and monitoring errors to flag anomalies, as shown in Figure 6 depicts a typical AE architecture.
Amir Basati et al. [93] presented a novel IDS for IoT networks utilizing Parallel Deep Auto-Encoders (PDAE). The proposed PDAE architecture combines dilated and conventional filters to capture local and surrounding information of input feature vectors. This method reduces the number of parameters and computational requirements, making it suitable for resource-constrained IoT devices. The model is evaluated on the KDD-Cup99, CICIDS2017, and UNSW-NB15 datasets, demonstrating superior accuracy compared to state-of-the-art algorithms, achieving high classification accuracy with significantly fewer parameters.
Khanam et al. [94] proposed a novel IDS using a Class-wise Focal Loss Variational Autoencoder (CFLVAE) to address data imbalance in IoT network traffic. The model generates new samples for minority attack classes and uses a Class-wise Focal Loss (CFL) to train a Variational Autoencoder (VAE), improving the quality and diversity of synthetic intrusion data. The balanced dataset is then used to train a deep neural network classifier.

Evaluated on the NSL-KDD dataset, the CFLVAE-DNN achieved an overall detection accuracy of 88.08%, a false positive rate of 3.77%, and high detection rates for low-frequency attacks (U2R: 79.25%, R2L: 67.5%), outperforming several state-of-the-art methods.
Aloul et al. [95] present a DL-based IDS designed for deployment on edge devices near IoT endpoints. The paper uses Adversarial Autoencoders (AAE) combined with the KNN algorithm to detect intrusions accurately. The model was trained on the NSL-KDD dataset, achieving an accuracy of 99.91% and an F1-score of 0.9990. The model was implemented on a Raspberry Pi 3B+ to evaluate real-world applicability, demonstrating an average inference time of 15.75 ms per packet, making it suitable for many IoT applications. The approach significantly reduces the computational load on central servers by performing IDS at the network edge.
Abdul Jabbar Siddiqui et al. [96] proposed methods to reduce the complexity of AE ensembles used for detecting network intrusions in IoT. The paper introduced four techniques for adaptively deactivating AE based on criteria-based and random approaches, which can be applied post-training or in-training. Extensive experiments on two realistic IoT-IDS datasets demonstrate that these methods significantly reduce training, retraining, and inference time costs while maintaining high detection performance.

### 4.0.4 Recurrent Neural Network (RNN)

RNNs are effective for IDS in IoT due to their ability to capture temporal dependencies in network traffic data. RNNs process data sequences, making them well-suited for detecting patterns associated with malicious activities over time. In IDS for IoT, RNNs can identify both short-term and long-term dependencies in network behavior, enhancing the detection accuracy of various cyberattacks. Implementing RNN-based IDS in a fog-cloud architecture helps reduce latency and computational load, enabling real-time IDS close to IoT devices. Figure 7 illustrates a typical RNN architecture, highlighting the structure that enables these capabilities.

Naeem et al. [97] proposed a fog-cloud-based IDS using RNN for IoT networks, address-ing challenges related to large IoT data volumes and computation requirements. The framework employs a dataset splitting method based on attack class and a feature selection process to reduce the dataset size by 90% , thus enhancing the DL models' efficiency. Two RNN algorithms, SimpleRNN and Bi-directional, are used for detecting intrusions.

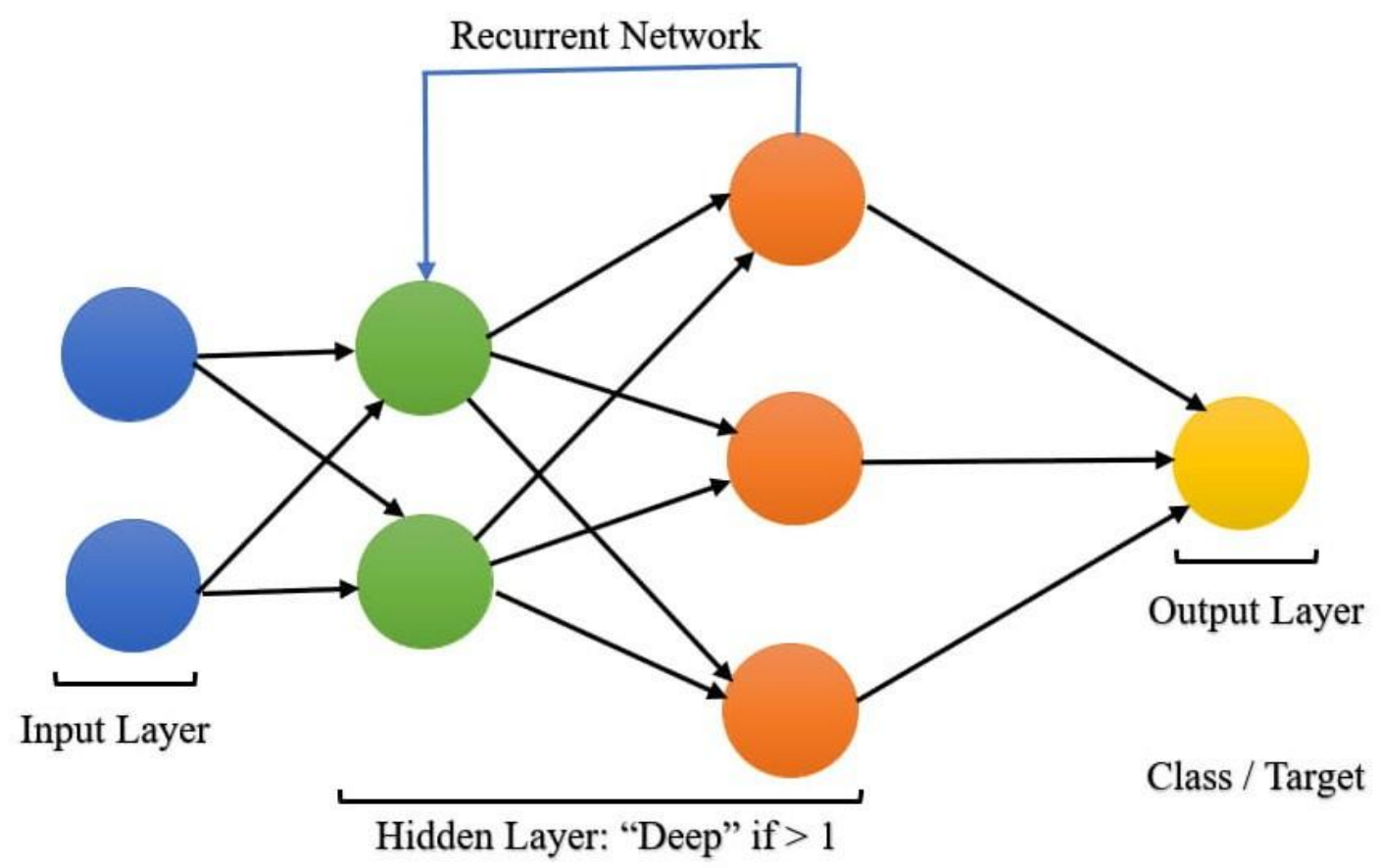

Figure 7: Typical RNN Architecture.

The BoT-IoT dataset evaluation demonstrates that the proposed models achieved higher recall rates and did not suffer from overfitting or underfitting issues.
Asma Belhadi et al. [98] proposed a novel framework for detecting group intrusions in IoT. The framework combines DL with a decomposition strategy, utilizing an Attention Segmental Recurrent Neural Network (ASRNN) to identify individual intrusions, which are then used to derive group outliers via a decomposition method. The approach was tested on two datasets, IDS 2018 and LUFlow, demonstrating superior performance in identifying group outliers compared to existing methods.
Ruijie Zhao et al. [99] proposed an efficient network IDS method for IoT using a lightweight deep neural network. The technique employs PCA for feature dimensionality reduction to lower the complexity of raw traffic features. It utilizes an expansion and compression structure, inverse residual structure, and channel shuffle operation for effective feature extraction with low computational cost. The proposed model incorporates a new loss function to handle the problem of uneven sample distribution. Evaluated on UNSW-NB15 and Bot-IoT datasets, the method demonstrated excellent classification performance with low model complexity and size, making it suitable for IoT.

#### 4.0.5 Deep Belief Network (DBN)

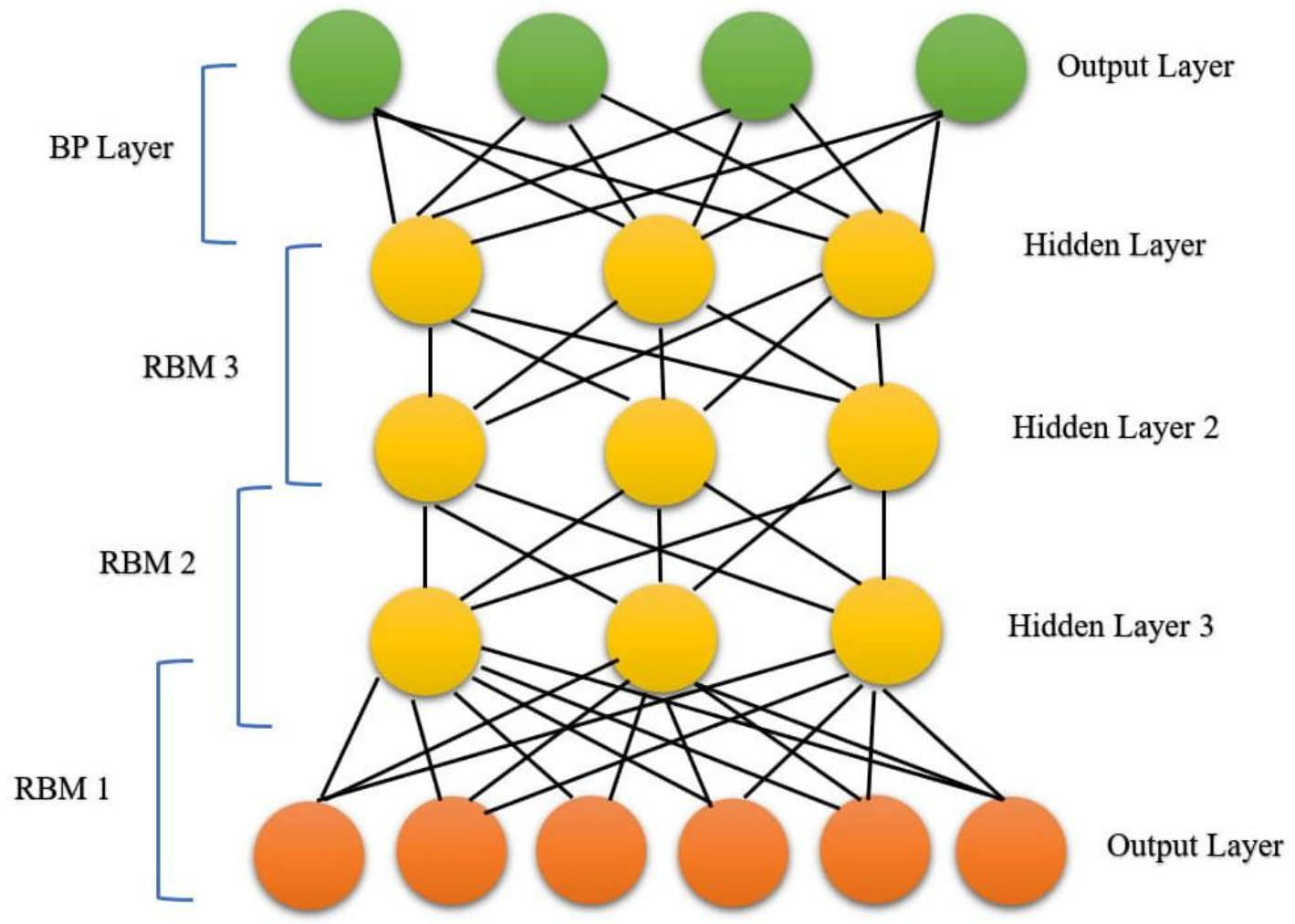

Figure 8: Typical DBN Architecture.

DBN is a type of DL model that excels in feature extraction and representation, making it highly suitable for IDS in IoT. DBNs are composed of multiple layers of Restricted Boltzmann Machines (RBMs) stacked together, enabling them to capture complex patterns in network traffic data. DBNs can effectively identify anomalous activities within IoT networks by leveraging unsupervised pre-training followed by supervised fine-tuning. Their hierarchical learning approach enhances detection accuracy, making DBNs a robust choice for securing IoT devices against cyber threats. Figure 8 shows a typical DBN architecture, demonstrating the layered structure that facilitates DL and pattern recognition.
Nagaraj Balakrishnan et al. [100] proposed using a DBN for improving IDS in IoT. The DBN, comprising multiple layers of RBMs, effectively classifies and identifies malicious activities by learning complex patterns in network traffic data. The paper demonstrates that their DBN-based IDS significantly enhances detection accuracy compared to traditional methods, with high performance across various types of attacks.

Rayeesa Malik et al. [101] proposed an advanced IDS leveraging a lightweight neural network approach tailored to IoT. The study introduces a novel methodology combining DBNs with an intelligent IDS mechanism that scrutinizes malicious activities within IoT networks. The implementation uses TensorFlow and evaluates the proposed DBN-Classifier on a subset of the ToN-IoT-Weather dataset. The findings reveal that the proposed model outperforms existing state-of-the-art techniques achieving an average accuracy of 86.3%.
Othmane Belarbi et al. [102] developed a DBN-based IDS to detect cyber-attacks in IoT. The proposed DBN model, trained and tested on the CICIDS2017 dataset, utilizes multiple class balancing techniques to address dataset imbalance issues. The study demonstrates that the DBN-based IDS outperforms a conventional Multi-Layer Perceptron (MLP) model and existing state-of-the-art methods, especially in detecting underrepresented attack classes. The DBN approach achieved significant performance improvements, with an F1-score increase from 0.87% to 0.94%, highlighting its effectiveness in enhancing detection capabilities in IoT.

Table 4: Comparison of DL Techniques in IDS for IoT

| **Feature** | **LSTM** | **CNN** | **AE** | **DBN** |
|---|---|---|---|---|
| **Type of Learning** | Supervised | Supervised | Unsupervised/Supervised | |
| **Classification Performance** | High | Very High | High | High |
| **Handling Temporal Data** | Excellent | Moderate | Poor | Moderate |
| **Feature Extraction** | Moderate | Excellent | Excellent | Excellent |
| **Anomaly Detection** | High | Moderate | Very High | High |
| **Scalability to Large Data** | Moderate | High | Very High | High |
| **Robustness to Noisy Data** | High | High | Moderate | High |
| **Training Complexity** | High | Moderate | Moderate | High |
| **Computational Complexity** | High | Moderate | Moderate | High |

Table 4 compares various DL techniques used in IDS for IoT, including LSTM, CNN, AE, and DBN. It highlights key features such as the type of learning, classification performance, ability to handle temporal data, feature extraction capabilities, anomaly detection effectiveness, scalability, robustness to noisy data, and training complexity. Each technique has distinct strengths, making them suitable for different aspects of IDS in IoT.
Table 5 compares ML-based and DL-based IDS in IoT. ML-based IDS are generally less complex and easier to implement and interpret, making them suitable for more straightforward problems and resource-constrained devices. They require less training data but heavily rely on manual feature engineering. In contrast, DL-based IDS excels in handling complex patterns and high-dimensional data through automatic feature extraction, achieving higher accuracy but at higher complexity and computational requirements.

Table 5: Comparison of ML-based and DL-based IDS in IoT

| Aspect | ML-based IDS | DL-based IDS |
|---|---|---|
| Complexity | Less complex, easier to implement and interpret. | Higher complexity requires more computational resources and expertise. |
| Training Data Requirements | Requires less training data, performs well with smaller datasets. | Requires large amounts of labeled data for high performance. |
| Feature Engineering | Relies on manual feature engineering, domain knowledge crucial. | Capable of automatic feature extraction, reducing the need for manual engineering. |
| Performance | Performs well on simpler, well-defined problems, struggles with complex patterns. | Excels at handling complex patterns and high-dimensional data, often with higher accuracy. |
| Adaptability | Less adaptability to new, unseen attack patterns, requiring retraining. | More adaptable to new and evolving attack patterns, can leverage continuous learning. |
| Scalability | Easier to scale on resource-constrained devices. | Challenging to scale on IoT devices due to higher computational and memory demands. |
| Interpretability | More interpretable, easier to understand and trust decisions. | Often considered a ”black box,” decisions are harder to interpret and explain. |
| Robustness to Adversarial Attacks | Less robust, can be easily fooled with crafted inputs. | Can be more robust with proper training, but still vulnerable to sophisticated attacks. |
| Deployment | Easier to deploy on resource-constrained IoT devices. | Requires more powerful hardware or offloading to edge/cloud computing. |

# 5 Challenges and Opportunities

ML-based IDS (including DL-based IDS) presents several significant challenges and pivotal opportunities for advancing IoT security.

## 5.1 Technical Challenges

**High False Alarm Rates:** One of the primary challenges in IDS is the high rate of false alarms. False positives, where normal network activities are misclassified as malicious, can overwhelm security analysts and reduce the system's credibility. This issue is especially pronounced in ML-based IDS due to the complexity and variability of network traffic. Future research should focus on developing advanced filtering mechanisms, adaptive learning models, and more robust validation processes to enhance the accuracy and reliability of IDS.

**Lack of Representative Datasets:** The availability of high-quality, representative datasets is crucial for training and evaluating IDS. Existing datasets, (e.g., KDD-CUP99 and NSL-KDD), often contain redundant data samples and limited attack categories, which do not reflect the current threat landscape. There is a need to create and utilize updated datasets that accurately represent real-world IoT network situations, including new attack classes and refined data samples such as those found in WUSTL-IIoT-2021.

**Real-time Intrusion Detection:** Handling network traffic's high speed and volume in real-time remains a significant challenge. IDS must process vast amounts of data quickly and efficiently to promptly detect and respond to threats. This necessitates optimizing algorithms for speed and efficiency, potentially through lightweight models, parallel processing, and hardware acceleration. Deploying lightweight IDS at the edge can enhance efficiency by processing data closer to the source, reducing latency, and distributing the computational load.

Alert Correlation: Effective alert correlation is essential for managing the large number of alerts generated by IDS. Current ML-based systems often struggle to aggregate and prioritize these alerts, leading to alert fatigue among security analysts. For example, an ML-based IDS might generate thousands of alerts for a single attack event, overwhelming analysts with redundant information. Future research should focus on developing intelligent correlation engines that can combine and rank alerts, improving the system's efficiency and effectiveness. Research by Julisch [103] has shown that clustering IDS alarms to identify and eliminate persistent root causes can significantly reduce the alarm load, as a few predominant root causes generally account for over 90% of the alarms.

**Proficiency in Handling Encrypted Network Traffic:** Effective alert correlation is essential for managing the large number of alerts generated by IDS. Current ML-based systems often struggle to aggregate and prioritize these alerts, leading to alert fatigue among security analysts. For example, an ML-based IDS might generate thousands of alerts for a single attack event, overwhelming analysts with redundant information. Future research should focus on developing intelligent correlation engines that can combine and rank alerts, improving the system's efficiency and effectiveness. For instance, Wang et al. [104] proposed a framework for network traffic classification that includes steps such as traffic dataset collection, feature selection, and algorithm selection to enhance the detection of encrypted malicious traffic. Additionally, the integration of deep packet inspection with ML algorithms can effectively analyze encrypted traffic patterns to detect anomalies and potential threats without compromising user privacy [105].

**Differentiating between Detecting False Data Injection and Attacks in ML-Based IDS:** ML-based IDS should differentiate between malicious attacks and benign anomalies, such as sensor malfunctions. For instance, a broken sensor might send incorrect data, triggering an alert. However, ML algorithms can be trained to recognize patterns of genuine attacks versus faulty sensor data. This capability ensures accurate threat detection while minimizing false positives, improving system reliability and efficiency. Recent research by Hu et al. [106] proposes a framework for detecting false data injection attacks in large-scale wireless sensor networks by leveraging spatiotemporal correlations in sensor data. This approach allows the system to differentiate between benign anomalies and malicious activities, thus enhancing detection accuracy and reducing false positives.

**Handling Intrusion Evasion Attacks:** Attackers continuously develop sophisticated

evasion techniques to bypass IDS. Designing adaptive models that can dynamically update and respond to these new evasion strategies is critical. Recent research by Han et al. [107] presented a study on evaluating and improving the adversarial robustness of ML-based IDS. Their work focuses on practical gray/black-box traffic-space adversarial attacks, proposing methods to enhance the robustness of ML-based IDS against sophisticated evasion strategies.
**Securing Industrial Control Systems:** Industrial control systems (ICS) are critical components of national infrastructure and are increasingly targeted by cyber-attacks. Developing specialized IDS tailored to the unique requirements of ICS, including resilience against targeted attacks and maintaining system availability, is a crucial area of research. For example, ML-based IDS can be trained on specific ICS network traffic patterns and operational behaviors to detect anomalies indicative of cyber-attacks. By continuously monitoring and learning from ICS data, these systems can identify and respond to threats (e.g., unauthorized access or data tampering), ensuring the continuous and secure operation of critical infrastructure [108].
**Adversarial Learning:** Adversarial attacks, where attackers manipulate input data to deceive ML models, pose a significant threat to IDS. Understanding and mitigating the impact of adversarial learning on IDS is an emerging area of research. This involves studying how adversarial samples affect ML models and developing robust defenses against such attacks.
**Energy Consumption and Performance Metrics:** The ML-based IDS should be rigorously tested under attack and non-attack scenarios to comprehensively measure energy consumption and other performance metrics. This dual evaluation ensures that the IDS is effective at detecting threats and efficient in terms of resource usage during normal operations. By assessing the IDS's performance in varied conditions, researchers can gain insights into its scalability, detection accuracy, and overall system impact, ensuring robustness and practicality in real-world deployments. Recent research by Tekin et al. [109] investigates the energy consumption of on-device ML models for IoT IDS applications. The study compares different ML algorithms regarding energy consumption during training and inference phases, highlighting the trade-offs between accuracy and energy efficiency for various approaches.

## 5.2 Generative AI in IoT security

LLMs as a representative element of Generative AI, benefiting from transformer-based architectures and vast amounts of data for pre-training, show promising results in generating and predicting various types of information, such as text and image [110]. In IoT security, the integration of Generative AI offers the potential for more secure, intelligent, and adaptive approaches to safeguarding the growing array of IoT devices and the network. The challenge lies in utilizing these cutting-edge technologies responsibly, ensuring that innovation is balanced with ethical considerations [111, 112].
The landscape of LLMs is diverse and rapidly evolving. Noteworthy models include BERT, introduced by Google in 2018, which utilizes a transformer-based architecture to convert data sequences, significantly enhancing Google search capabilities [113] [114]. Claude, developed by Anthropic, focuses on constitutional AI, producing outputs that are helpful, harmless, and accurate [115] [116]. Other prominent models include Ernie by Baidu, renowned for its proficiency in Mandarin and extensive user base [117]. The Technology Innovation Institute's Falcon 40B/180B, an open-source model available on Amazon

SageMaker, is another key player [118]. OpenAI's GPT series, particularly influential with GPT-3's introduction of 175 billion parameters in 2020, has been pivotal in AI development [119] [120]. GPT-3.5, with fewer parameters, powers ChatGPT and integrates with Bing search, while the latest GPT-4, released in 2023, is a multimodal model capable of processing both language and images [121]. Microsoft's Orca demonstrates efficiency with fewer parameters, and Google's Palm specializes in complex reasoning tasks across various domains [122] [123]. Smaller, specialized models like Microsoft's Phi-1 prioritize data quality over quantity [124] [125]. Stability AI's StableLM series emphasizes transparency and accessibility. Finally, Vicuna 33B, a fine-tuned version of LLaMA, is a notable open-source model with a smaller parameter count but effective capabilities [126] [127].

### 5.2.1 Vulnerability Management

LLMs are adept at identifying and managing vulnerabilities in IoT devices by analyzing firmware, software updates, and configuration settings. By detecting potential security flaws and suggesting patches or configuration changes, LLMs help maintain the security and integrity of IoT systems.

### 5.2.2 Predictive Maintenance

LLMs can analyze historical performance data to predict when IoT devices are likely to fail or become vulnerable to attacks. This capability allows for proactive maintenance and updates, preventing issues before they arise and ensuring IoT systems' continuous, secure operation.

### 5.2.3 Advanced Threat Detection and Classification

To enhance real-time detection and classification of attacks, LLMs fine-tuned on datasets such as CICIDS2017 can be utilized. These models can accurately identify specific attack types, including DDoS and port scanning. For instance, upon detecting a DDoS attack pattern, the LLMs can immediately classify it, thereby significantly improving detection accuracy and responsiveness.

### 5.2.4 Intelligent Security Policy Recommendations

To enhance the deployment of LLMs in IoT security, these models can be used to generate context-aware security policy recommendations upon detecting various attacks. For example, when a DDoS attack is identified, the LLMs can provide actionable advice such as: ”We have activated the firewall and blocked IP address X”. This approach ensures timely and precise defensive measures, enhancing the overall security posture of IoT systems.

### 5.2.5 Cyber Threat Detection with Generative AI

Generative AI can revolutionize IoT security by analyzing vast amounts of unstructured data, such as device logs and network traffic, to detect anomalies indicative of cyber-attacks. For example, an LLM can identify unusual patterns, such as sudden spikes in outbound data, signaling potential data exfiltration [128]. This proactive approach enhances threat detection and reduces false positives, providing robust protection for IoT networks.

### 5.2.6 Incident Response Automation

LLMs can streamline the response to security incidents by offering real-time recommendations or executing predefined actions. This automation helps quickly isolate compromised devices, log incidents, and alert security teams with detailed reports and remediation steps, thereby mitigating threats efficiently.

### 5.2.7 Personalized User Recommendations for Cyber Threat Mitigation

Based on the analysis of security threats, LLMs can provide personalized recommendations to users for mitigating potential cyber threats. LLMs can suggest specific actions to enhance security by evaluating user behavior and device usage patterns. For example, suppose an LLM detects unusual activity on a user's smart home network. In that case, it can recommend changing passwords, enabling multi-factor authentication, updating device firmware, or avoiding specific actions that may increase vulnerability. These personalized recommendations help users safeguard their IoT devices and networks against potential threats.

### 5.2.8 Penetration testing

Generative AI can significantly improve penetration testing in IoT networks by automating and optimizing various aspects of the process. These AI tools can create realistic phishing emails and social engineering content, simulate advanced cyber-attacks to evaluate network defenses, and predict potential vulnerabilities by analyzing code or system configurations. Additionally, they can generate custom scripts for testing applications and networks and use natural language processing to interpret and analyze penetration test results, providing insights and recommendations for enhancing security. LLMs can also be fine-tuned to the latest cybersecurity trends and exploits, ensuring current penetration tests cover a broad range of potential threats [129].

### 5.2.9 IoT security

Generative AI significantly enhances penetration testing in IoT networks by automating and optimizing various processes. These AI tools can craft realistic phishing emails and social engineering scenarios, simulate complex cyber-attacks to test network defenses, and predict potential vulnerabilities by analyzing code or system configurations. They also assist in generating custom scripts for testing applications and networks and use natural language processing to interpret and analyze the results of penetration tests, providing insights and recommendations for improving security [129,130]. For example, an LLM can analyze IoT device logs to identify unusual access patterns that may indicate an attempted breach. If a sudden spike in outbound data is detected, the model can flag it as a potential data exfiltration attempt, allowing security teams to respond swiftly. Moreover, LLMs can stay updated with the latest cybersecurity trends and exploits, ensuring comprehensive and current penetration tests. This proactive approach enhances threat detection and reduces false positives, providing robust protection for IoT networks.

## 5.3 Ethical Considerations and Privacy

The deployment of ML-based IDS in IoT environments raises several ethical and privacy concerns that must be carefully addressed to ensure user data protection and the eth-ical use of technology. One primary concern is collecting and processing large volumes of sensitive data. For instance, smart home devices continuously collect data about the inhabitants' behaviors and routines. An ML-based IDS monitoring this data for security must ensure the collected data is anonymized and securely stored to prevent unauthorized access and misuse.

Another significant ethical consideration is the potential for bias in ML models. If an IDS is trained on biased data, it may disproportionately target or ignore specific types of devices or behaviors, leading to unfair treatment of users. For example, an IDS trained primarily on data from European smart homes might not perform as well in other cultural contexts, leading to false positives or negatives. It is crucial to use diverse datasets during the training phase to mitigate bias and ensure fair treatment across different user groups and environments.

Privacy concerns are also paramount when considering the potential for these systems to be used for surveillance beyond their intended security functions. For instance, a DL-based IDS with advanced anomaly detection capabilities could inadvertently monitor and analyze personal habits and movements within a smart city, infringing on individuals' privacy rights. To address this, it is essential to implement strict data governance policies, ensuring that data is used solely for security purposes and not for unwarranted surveillance. Furthermore, transparency and user consent are critical ethical considerations. Users should be fully informed about what data is being collected, how it is being used, and the purpose of the IDS. For example, in a healthcare IoT system, patients must be aware of how their health data is monitored and protected by IDS, and they should consent to its usage. Clear privacy policies and obtaining explicit user consent can help maintain trust and ensure ethical compliance.

# 6 Discussion

Addressing the challenges in ML-based IDS for IoT requires detailed and multifaceted approaches. Advanced filtering mechanisms, ensemble learning, and post-processing algorithms can be employed to reduce high false alarm rates. These mechanisms dynamically adjust detection thresholds, combine multiple models, and analyze alerts to minimize false positives. The lack of representative datasets can be overcome by generating synthetic data using techniques (e.g., GANs), promoting collaborative data sharing with standardized protocols, and continuously updating datasets with new attack types. Real-time IDS can be enhanced through edge computing, stream processing frameworks such as Apache Kafka and Flink, and hardware accelerators, e.g., GPUs and FPGAs. Effective alert correlation involves developing intelligent correlation engines that aggregate and prioritize alerts, incorporating contextual analysis, and implementing automated response systems. Encrypted network traffic can be handled through deep packet inspection techniques, secure encryption methods, and ML algorithms that analyze encrypted traffic patterns. Differentiating malicious attacks from benign anomalies requires context-aware detection, hybrid models combining multiple detection techniques, and continuous learning models that update with new patterns. Addressing intrusion evasion attacks involves adversarial training, behavioral analysis, and adaptive models that update detection strategies in real-time. Securing industrial control systems necessitates tailored IDS models for specific protocols, redundancy, resilience measures, and regular security audits. Imbalanced classes in datasets can be managed with over-sampling, under-sampling, cost-sensitive learning, and anomaly detection models. Handling concept drift involves online learning, periodic retraining, and ensemble methods combining models trained on different peri-ods. Mitigating adversarial learning impacts requires robust training techniques, defense mechanisms like input sanitization, and continuous monitoring of IDS models.

Table 6: Comprehensive Solutions for Ethical and Privacy Concerns in ML-Based IDS Deployment in IoT Environments

| Issue | Solution | Implementation Steps |
|---|---|---|
| **Data Anonymization and Security** | **Data Anonymization** | - Utilize k-anonymity, differential privacy, and data masking to anonymize data. Conduct regular audits to verify the effectiveness of anonymization processes. |
| **Bias Mitigation in ML Models** | **Diverse Training Datasets** | - Ensure the collection of diverse and representative datasets from various sources and demographics. To mitigate biases, apply bias detection and correction techniques, such as re-weighting and re-sampling. |
| | **Regular Audits and Testing** | - Use fairness testing tools and frameworks to evaluate model performance continuously. Conduct periodic audits to identify and correct biases, ensuring models remain fair and accurate across different contexts. |
| **Prevention of Unwarranted Surveillance** | **Strict Data Governance Policies** | - Develop and enforce clear data governance frameworks that specify permissible data uses. Implement access controls and monitoring mechanisms to ensure compliance with governance policies. |
| | **Anomaly Detection Limits** | - Define clear criteria for security-related anomalies that respect personal privacy. Regularly review and update anomaly detection criteria to reflect ethical considerations and prevent overreach. |
| **Transparency and User Consent** | **Clear Privacy Policies** | - Develop comprehensive and user-friendly privacy notices and documentation. Ensure that privacy policies are easily accessible and understandable to all users. |
| | **Explicit User Consent** | - Implement opt-in consent mechanisms to obtain explicit user consent. Provide users with options to manage their consent preferences, ensuring informed participation. |

Finally, reducing energy consumption and improving performance metrics involve developing efficient algorithms, leveraging edge computing to distribute processing loads, and conducting comprehensive evaluations of IDS models to measure energy consumption, latency, and other performance metrics under various scenarios.

Integrating Generative AI into IoT security solutions offers significant potential to address some challenges more effectively. For example, LLMs can generate realistic synthetic datasets to improve training and testing by creating varied and complex attack scenarios, which traditional methods may not cover. An example is using LLMs to create synthetic data that mirrors the complexity of real-world IoT network traffic, enhancing the robustness of IDS. LLMs can also improve alert correlation by analyzing and contextualizing alerts, reducing false positives. For instance, using BERT, LLMs can process and interpret the context of network activity logs to differentiate between genuine threats and benign anomalies more accurately. This helps prioritize alerts and reduces the workload of security analysts.

In handling encrypted traffic, LLMs, like GPT-4, can detect patterns and anomalies without decrypting the data, thereby maintaining privacy. For example, by analyzing metadata and traffic flow characteristics, LLMs can be trained to recognize traffic behavior patterns indicative of malicious activity, even when the payload is encrypted. Additionally, LLMs can continuously learn from new data to adapt to emerging threats. For instance, an LLMs-based system can be updated with the latest threat intelligence and automatically adjust its detection algorithms to account for new types of attacks, ensuring that the IDS remains effective over time.

Integrating LLMs into IDS frameworks presents a promising direction for future research. This aim is to create more robust, adaptive, and efficient IoT security solutions. By leveraging LLMs' advanced natural language processing capabilities, we can enhance IDS's effectiveness and reliability in protecting complex and dynamic IoT environments.

# 7 Conclusions

In the rapidly expanding IoT landscape, securing interconnected devices is critical. This chapter explored the integration of advanced ML and DL techniques into IDS to combat complex cyber threats. Traditional IDS methods often fall short due to the vast, diverse data from IoT devices and sophisticated attacks. ML offers significant enhancements in detection, adaptation, and mitigation. We classified IDS methodologies and examined the strengths and weaknesses of various ML techniques. Key challenges discussed included dataset quality, model bias, overfitting, computational efficiency, and scalability. The importance of ethical considerations and privacy was also highlighted. In conclusion, integrating ML and DL into IDS promises to enhance IoT security by adapting to new threats. Ongoing research and development are essential to protecting the interconnected world and ensuring security and privacy for all users. This comprehensive analysis will guide future research and implementation of robust, adaptive, and intelligent IDS solutions for IoT ecosystems.

# Author Biographies

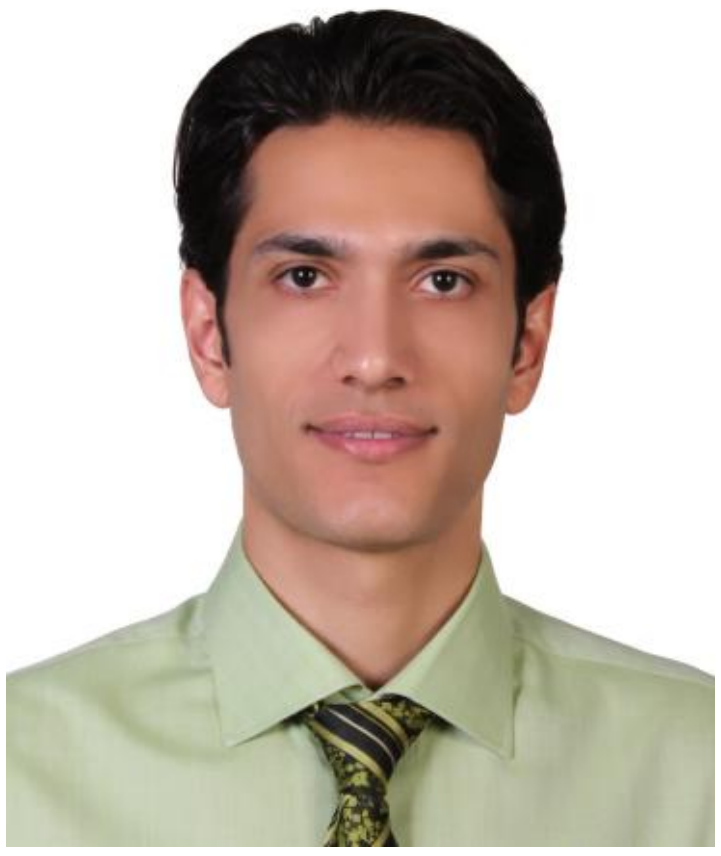

**Saeid Jamshidi** holds a Bachelor's degree in Software Engineering and a Master's degree in Computer Engineering (Information Systems), with a specialization in Communications and the Internet of Things (IoT). He completed his Ph.D. in Cybersecurity at École Polytechnique de Montréal, where his research focused on AI-driven cybersecurity. His work explored the application of deep reinforcement learning (DRL), machine learning (ML), and Green AI in intrusion detection and IoT security systems. He is pursuing a postdoctoral fellowship at École Polytechnique de Montréal, in collaboration with Google and the Interdisciplinary Research Centre on Cybersecurity (IMC), where he is continuing to advance research in intelligent, sustainable cybersecurity and secure IoT infrastructures.

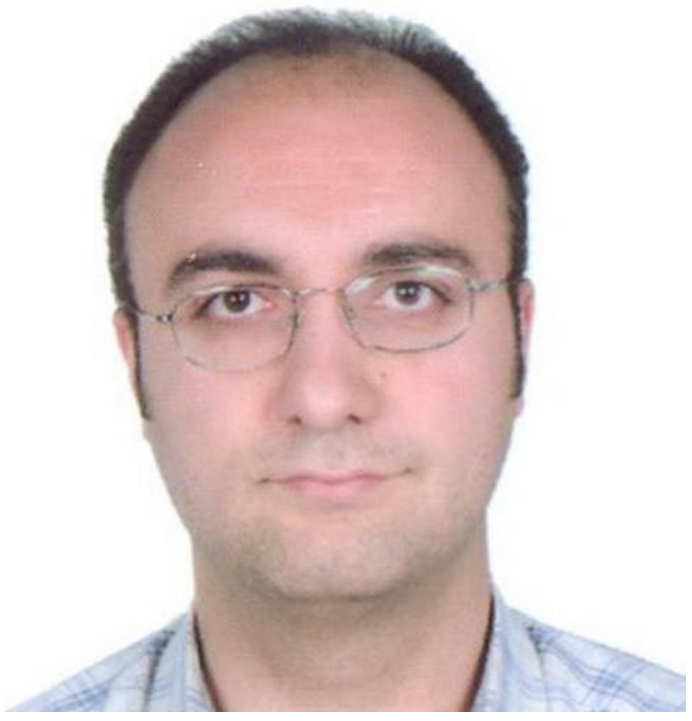

**Amin Nikanjam** is a Staff Researcher at Huawei Canada. He is investigating how Software Engineering practices—such as testing and fault localization—can be leveraged in Machine Learning Software Systems and how Machine Learning techniques can be applied to safety-critical systems regarding reliability, robustness, and explainability. He received his Master's and Ph.D. in Artificial Intelligence from Iran University of Science and Technology and his Bachelor's in Software Engineering from the University of Isfahan. He was formerly a Research Associate at Polytechnique Montréal, an Invited Researcher at the University of Montréal, and an Assistant Professor at K. N. Toosi University of Technology, Iran. His research interests include Software Engineering for Machine Learning, Machine Learning Systems Engineering, Large Language Models for SE, and Multi-Agent Systems.

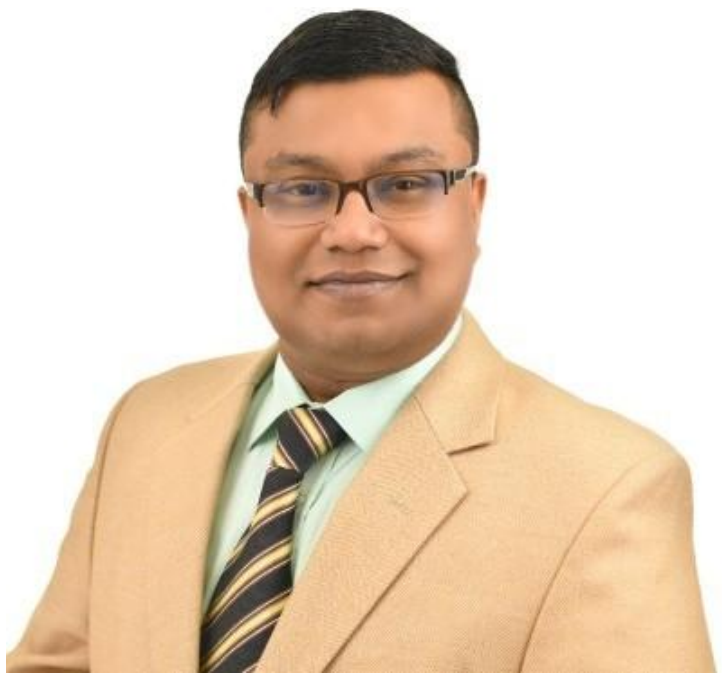

**Nafi Kawser Wazed** is a Research Student at the Department of Computer Science at École Polytechnique de Montréal. His research interests include Software Engineering, Source Code Analysis, Cross-Language Software Development and Maintenance, Wireless Ad-hoc Networking, and Network Security. He has published several papers in international conferences and journals. He was a Ph.D. student at the University of Saskatchewan and an M.Sc. student at the University of Ottawa.

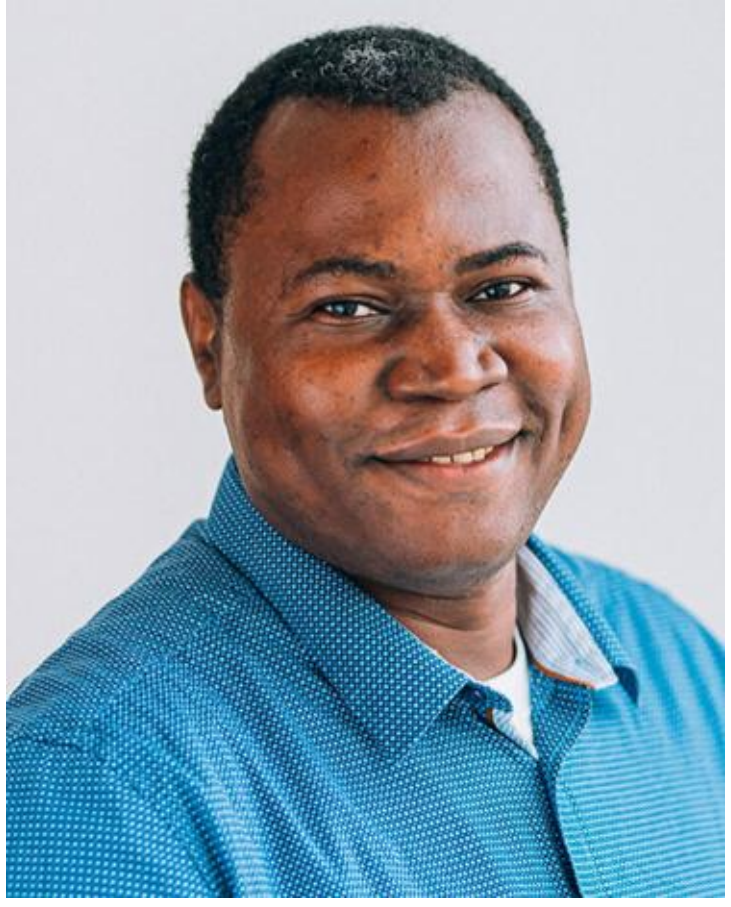

**Foutse Khomh** is a Full Professor of Software Engineering at Polytechnique Montréal, Canada CIFAR AI Chair on Trustworthy ML Software Systems, and FRQ-IVADO Research Chair on Software Quality Assurance for ML Applications. He received a Ph.D. in Software Engineering from the University of Montreal in 2011, with the Award of Excellence. He also received a CS-Can/Info-Can Outstanding Young Computer Science Researcher Prize for 2019. His research interests include software maintenance and evolution, ML systems engineering, cloud engineering, and dependable and trustworthy ML/AI. His work has received four ten-year Most Influential Paper (MIP) Awards, and six Best/Distinguished Paper Awards. He has served on the program committees of several international conferences including ICSE, FSE, ICSM(E), SANER, and MSR, and has reviewed for top international journals such as EMSE, TSC, TPAMI, TSE, and TOSEM. He also served on the steering committee of SANER (chair), MSR, PROMISE, ICPC (chair), and ICSME (vice-chair). He initiated and co-organized the Software Engineering for ML Applications (SEMLA) symposium and the RELENG (Release Engineering) workshop series. He is co-founder of the NSERC CREATE SE4AI: A Training Program on the Development, Deployment, and Servicing of Artificial Intelligence-based Software Systems, and one of the Principal Investigators of the DEpendable Explainable Learning (DEEL) project. He is on the editorial board of multiple international software engineering journals and is a Senior Member of IEEE.